EUROPEAN ORGANIZATION FOR NUCLEAR RESEARCH

# First Measurement of Chiral Dynamics in $\pi^- \gamma \to \pi^- \pi^- \pi^+$

The COMPASS Collaboration

**Abstract**

The COMPASS collaboration at CERN has investigated the $\pi^- \gamma \to \pi^- \pi^- \pi^+$ reaction at center-of-momentum energy below five pion masses, $\sqrt{s} < 5m_\pi$, embedded in the Primakoff reaction of 190 GeV pions impinging on a lead target. Exchange of quasi-real photons is selected by isolating the sharp Coulomb peak observed at smallest momentum transfers, $t' < 0.001$ GeV$^2/c^2$. Using partial-wave analysis techniques, the scattering intensity of Coulomb production described in terms of chiral dynamics and its dependence on the $3\pi$-invariant mass $m_{3\pi} = \sqrt{s}$ were extracted. The absolute cross section was determined in seven bins of $\sqrt{s}$ with an overall precision of 20%. At leading order, the result is found to be in good agreement with the prediction of chiral perturbation theory over the whole energy range investigated.

*(to be submitted to Physical Review Letters)*

## The COMPASS Collaboration

M.G. Alekseev[28], V.Yu. Alexakhin[7], Yu. Alexandrov[15], G.D. Alexeev[7], A. Amoroso[27], A. Austregesilo[10,17], B. Badełek[30], F. Balestra[27], J. Barth[4], G. Baum[1], Y. Bedfer[22], J. Bernhard[13], R. Bertini[27], M. Bettinelli[16], K. Bicker[10,17], R. Birsa[24], J. Bisplinghoff[3], P. Bordalo[12,a], F. Bradamante[25], A. Bravar[24], A. Bressan[25], E. Burtin[22], D. Chaberny[13], M. Chiosso[27], S.U. Chung[17,b], A. Cicuttin[26], M.L. Crespo[26], S. Dalla Torre[24], S. Das[6], S.S. Dasgupta[6], O.Yu. Denisov[10,28], L. Dhara[6], S.V. Donskov[21], N. Doshita[2,32], V. Duic[25], W. Dünnweber[16], M. Dziewiecki[30], A. Efremov[7], P.D. Eversheim[3], W. Eyrich[8], M. Faessler[16], A. Ferrero[22], A. Filin[21], M. Finger[19], M. Finger jr.[7], H. Fischer[9], C. Franco[12], N. du Fresne von Hohenesche[10,13], J.M. Friedrich[17], R. Garfagnini[27], F. Gautheron[2], O.P. Gavrichtchouk[7], R. Gazda[30], S. Gerassimov[15,17], R. Geyer[16], M. Giorgi[25], I. Gnesi[27], B. Gobbo[24], S. Goertz[2,4], S. Grabmüller[17], A. Grasso[27], B. Grube[17], R. Gushterski[7], A. Guskov[7], F. Haas[17], D. von Harrach[13], T. Hasegawa[14], F.H. Heinsius[9], F. Herrmann[9], C. Heß[2], F. Hinterberger[3], N. Horikawa[18,c], Ch. Höppner[17], N. d'Hose[22], S. Huber[17], S. Ishimoto[18,d], O. Ivanov[7], Yu. Ivanshin[7], T. Iwata[32], R. Jahn[3], P. Jasinski[13], G. Jegou[22], R. Joosten[3], E. Kabuß[13], D. Kang[9], B. Ketzer[17], G.V. Khaustov[21], Yu.A. Khokhlov[21], Yu. Kisselev[2], F. Klein[4], K. Klimaszewski[30], S. Koblitz[13], J.H. Koivuniemi[2], V.N. Kolosov[21], K. Kondo[2,32], K. Königsmann[9], I. Konorov[15,17], V.F. Konstantinov[21], A. Korzenev[22,e], A.M. Kotzinian[27], O. Kouznetsov[7,22], M. Krämer[17], Z.V. Kroumchtein[7], F. Kunne[22], K. Kurek[30], L. Lauser[9], A.A. Lednev[21], A. Lehmann[8], S. Levorato[25], J. Lichtenstadt[23], A. Maggiora[28], A. Magnon[22], N. Makke[22], G.K. Mallot[10], A. Mann[17], C. Marchand[22], A. Martin[25], J. Marzec[31], F. Massmann[3], T. Matsuda[14], W. Meyer[2], T. Michigami[32], Yu.V. Mikhailov[21], M.A. Moinester[23], A. Morreale[22], A. Mutter[9,13], A. Nagaytsev[7], T. Nagel[17], F. Nerling[9], S. Neubert[17], D. Neyret[22], V.I. Nikolaenko[21], W.-D. Nowak[9], A.S. Nunes[12], A.G. Olshevsky[7], M. Ostrick[13], A. Padee[31], R. Panknin[4], D. Panzieri[29], B. Parsamyan[27], S. Paul[17], E. Perevalova[7], G. Pesaro[25], D.V. Peshekhonov[7], G. Piragino[27], S. Platchkov[22], J. Pochodzalla[13], J. Polak[11,25], V.A. Polyakov[21], G. Pontecorvo[7], J. Pretz[4], C. Quintans[12], J.-F. Rajotte[16], S. Ramos[12,a], V. Rapatsky[7], G. Reicherz[2], A. Richter[8], E. Rocco[10], E. Rondio[30], D.I. Ryabchikov[21], V.D. Samoylenko[21], A. Sandacz[30], M.G. Sapozhnikov[7], S. Sarkar[6], I.A. Savin[7], G. Sbrizzai[25], P. Schiavon[25], C. Schill[9], T. Schlüter[16], L. Schmitt[17,f], K. Schönning[10], S. Schopferer[9], W. Schröder[8], O.Yu. Shevchenko[7], H.-W. Siebert[13], L. Silva[12], L. Sinha[6], A.N. Sissakian[7,*], M. Slunecka[7], G.I. Smirnov[7], S. Sosio[27], F. Sozzi[25], A. Srnka[5], M. Stolarski[12], M. Sulc[11], R. Sulej[30], P. Sznajder[30], S. Takekawa[25], S. Tessaro[24], F. Tessarotto[24], A. Teufel[8], L.G. Tkatchev[7], S. Uhl[17], I. Uman[16], M. Vandenbroucke[22], M. Virius[20], N.V. Vlassov[7], R. Windmolders[4], W. Wiślicki[30], H. Wollny[9,22], K. Zaremba[31], M. Zavertyaev[15], E. Zemlyanichkina[7], M. Ziembicki[31], N. Zhuravlev[7], and A. Zvyagin[16]

[1] Universität Bielefeld, Fakultät für Physik, 33501 Bielefeld, Germany[g]
[2] Universität Bochum, Institut für Experimentalphysik, 44780 Bochum, Germany[g]
[3] Universität Bonn, Helmholtz-Institut für Strahlen- und Kernphysik, 53115 Bonn, Germany[g]
[4] Universität Bonn, Physikalisches Institut, 53115 Bonn, Germany[g]
[5] Institute of Scientific Instruments, AS CR, 61264 Brno, Czech Republic[h]
[6] Matrivani Institute of Experimental Research & Education, Calcutta-700 030, India[i]
[7] Joint Institute for Nuclear Research, 141980 Dubna, Moscow region, Russia[j]
[8] Universität Erlangen–Nürnberg, Physikalisches Institut, 91054 Erlangen, Germany[g]
[9] Universität Freiburg, Physikalisches Institut, 79104 Freiburg, Germany[g]
[10] CERN, 1211 Geneva 23, Switzerland
[11] Technical University in Liberec, 46117 Liberec, Czech Republic[h]
[12] LIP, 1000-149 Lisbon, Portugal[k]
[13] Universität Mainz, Institut für Kernphysik, 55099 Mainz, Germany[g]
[14] University of Miyazaki, Miyazaki 889-2192, Japan[l]
[15] Lebedev Physical Institute, 119991 Moscow, Russia

[16] Ludwig-Maximilians-Universität München, Department für Physik, 80799 Munich, Germany[f,l)]
[17] Technische Universität München, Physik-Department, 85748 Garching, Germany[f,l)]
[18] Nagoya University, 464 Nagoya, Japan[l]
[19] Charles University in Prague, Faculty of Mathematics and Physics, 18000 Prague, Czech Republic[h]
[20] Czech Technical University in Prague, 16636 Prague, Czech Republic[h]
[21] State Research Center of the Russian Federation, Institute for High Energy Physics, 142281 Protvino, Russia
[22] CEA IRFU/SPhN Saclay, 91191 Gif-sur-Yvette, France
[23] Tel Aviv University, School of Physics and Astronomy, 69978 Tel Aviv, Israel[n]
[24] Trieste Section of INFN, 34127 Trieste, Italy
[25] University of Trieste, Department of Physics and Trieste Section of INFN, 34127 Trieste, Italy
[26] Abdus Salam ICTP and Trieste Section of INFN, 34127 Trieste, Italy
[27] University of Turin, Department of Physics and Torino Section of INFN, 10125 Turin, Italy
[28] Torino Section of INFN, 10125 Turin, Italy
[29] University of Eastern Piedmont, 1500 Alessandria, and Torino Section of INFN, 10125 Turin, Italy
[30] National Centre for Nuclear Research and University of Warsaw, 00-681 Warsaw, Poland[o]
[31] Warsaw University of Technology, Institute of Radioelectronics, 00-665 Warsaw, Poland[o]
[32] Yamagata University, Yamagata, 992-8510 Japan[l]

[a] Also at IST, Universidade Técnica de Lisboa, Lisbon, Portugal
[b] Supported by IAS-TUM, DFG and Humboldt foundation
[c] Also at Chubu University, Kasugai, Aichi, 487-8501 Japan[l]
[d] Also at KEK, 1-1 Oho, Tsukuba, Ibaraki, 305-0801 Japan
[e] On leave of absence from JINR Dubna
[f] Also at GSI mbH, Planckstr. 1, D-64291 Darmstadt, Germany
[g] Supported by the German Bundesministerium für Bildung und Forschung
[h] Suppported by Czech Republic MEYS grants ME492 and LA242
[i] Supported by SAIL (CSR), Govt. of India
[j] Supported by CERN-RFBR grants 08-02-91009
[k] Supported by the Portuguese FCT - Fundação para a Ciência e Tecnologia, COMPETE and QREN, grants CERN/FP/83542/2008, CERN/FP/109323/2009 and CERN/FP/116376/2010
[l] Supported by the MEXT and the JSPS under the Grants No.18002006, No.20540299 and No.18540281; Daiko Foundation and Yamada Foundation
[m] Supported by the DFG cluster of excellence 'Origin and Structure of the Universe' (www.universe-cluster.de)
[n] Supported by the Israel Science Foundation, founded by the Israel Academy of Sciences and Humanities
[o] Supported by Ministry of Science and Higher Education grant 41/N-CERN/2007/0
[*] Deceased

For hadronic interactions at low energy, quantum chromodynamics (QCD) can be formulated in terms of an effective field theory resulting from the systematic treatment of chiral symmetry and its breaking pattern, called chiral perturbation theory (ChPT). In this approach, the pions $(\pi^+, \pi^0, \pi^-)$ are identified as the Goldstone bosons associated with the spontaneous breaking of the chiral symmetry. Properties and interactions of the pions provide the most rigorous test of ChPT as the correct low-energy representation of QCD.

ChPT has accurately predicted the dynamics of low-energy $\pi\pi$ scattering as measured in various kaon decay experiments, see Refs. in [1]. High-precision calculations of the pion polarizability in the $\pi\gamma \to \pi\gamma$ reaction have been performed, which however are at large variance with the existing experimental determinations [2].

In view of this situation it is of great interest to study other low-energy reactions involving pions and photons, apart from measuring the pion polarizabilities with improved accuracy as proposed at COMPASS-II [3]. The $\pi\gamma \to 3\pi$ reaction examined here allows for testing the chiral dynamics of pions in the region near threshold where calculations are most reliable [1].

COMPASS at the CERN Super Proton Synchrotron is a large-acceptance, high-precision spectrometer [4], which is particularly well suited for investigations of high-energy reactions of particles at low to intermediate momentum transfer to fixed targets. The apparatus acceptance fully covers the phase space for reaction products emerging in forward direction. At very low momentum transfer in the process $\pi^-\mathrm{Pb} \to X^-\mathrm{Pb}$, photon exchange becomes important and competes with strong interaction processes like diffractive excitation via Pomeron exchange. The separation of these two contributions is an important aspect of this Letter.

Primakoff reactions are peripheral hadronic reactions in the quasi-real photon field surrounding a heavy nucleus [5]. The $\pi$-nucleus cross section can be connected to the $\pi\gamma$ cross section using the well-known equivalent photon approximation (EPA) [6]

$$\frac{d\sigma_{\mathrm{Pb}}^{\mathrm{EPA}}}{ds\,dt'\,d\Phi_n} = \frac{Z^2\alpha}{\pi(s-m_\pi^2)} \cdot F_{\mathrm{eff}}^2(t') \cdot \frac{t'}{(t'+|t|_{\mathrm{min}})^2} \cdot \frac{d\sigma_\gamma}{d\Phi_n} \tag{1}$$

Here, the cross section for the process $\pi^-\mathrm{Pb} \to X^-\mathrm{Pb}$ is factorized into the quasi-real photon density provided by the nucleus of charge $Z$, and $\sigma_\gamma$, the cross section for the embedded $\pi^-\gamma \to X^-$ scattering of a pion and a real photon. The mass of the charged pion is denoted $m_\pi$, and $d\Phi_n$ is the $n$-particle phase-space element of the final-state system $X^-$. From the 4-momentum transfer squared $t = (p_\pi - p_X)^2$, the positive quantity $t' = |t| - |t|_{\mathrm{min}}$ is derived with $|t|_{\mathrm{min}} = (s-m_\pi^2)^2/(4E_{\mathrm{beam}}^2)$ for a given final-state mass $m_X = \sqrt{s}$, where $s = (p_\pi + p_\gamma)^2$ is the squared centre-of-momentum energy. The lead form factor is approximated by the equivalent sharp radius formula $F_{\mathrm{eff}}(t') = j_1(r\sqrt{t'})$ with $r = 6.84\,\mathrm{fm}$.

For the specific reaction studied in this Letter with $X^- \to \pi^-\pi^-\pi^+$, the Feynman graphs for $\sigma_{\mathrm{Pb}}^{\mathrm{EPA}}$ are depicted in Fig. 1(a). The Dalitz plot distributions given in Fig. 1(b), which represent a 2-dimensional projection of the 5-dimensional phase space $\Phi_3$, show that the momenta of the 3 outgoing pions are far from being uniformly distributed in the $m_{3\pi}$ region near threshold.

The data presented in the following were recorded in the year 2004 with a 190 GeV hadron beam (at the target composed of 96.8% $\pi^-$, 2.4% $K^-$, 0.8% $\bar{p}$) interacting with a 3 mm lead target. The trigger selected events with at least two outgoing charged particles at scattering angles less than about 50 mrad. In the data analysis, exactly three outgoing charged particles, assumed to be pions, were required to form with the incoming beam particle a vertex that is consistent with an interaction within the target volume. Neglecting the tiny target nucleus recoil at low $t'$, their energy sum $E_{3\pi}$ equals the beam energy for the exclusive $\pi^-\mathrm{Pb} \to \pi^-\pi^-\pi^+\mathrm{Pb}$ reaction. Taking into account the spread of the beam energy, $\sigma_{\mathrm{beam}} \approx 1.3$ GeV, $E_{3\pi}$ is required to lie within $\pm 4$ GeV around the mean beam energy. Figure 2 shows the $m_{3\pi}$ distribution with the requirement $t' < 0.001\,\mathrm{GeV}^2/c^2$, *i.e.* the Primakoff-$t'$ region. The main

(a)
(b)

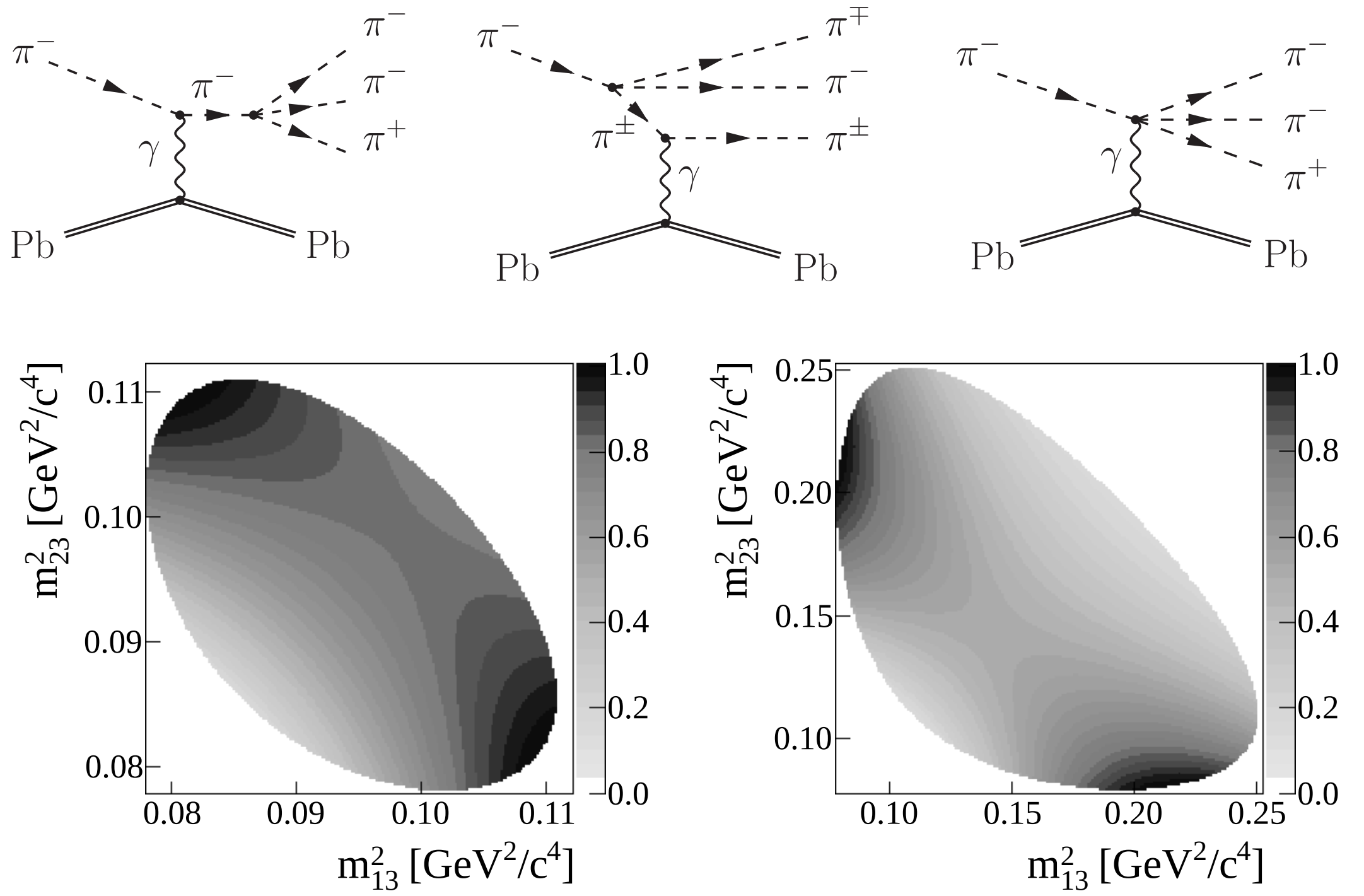

**Fig. 1:** (a) Leading-order, *i.e.* tree-level ChPT diagrams for $\pi^-\gamma \to \pi^-\pi^-\pi^+$, embedded in the Primakoff reaction $\pi^-\mathrm{Pb} \to \pi^-\pi^-\pi^+\mathrm{Pb}$. (b) Dalitz plots corresponding to the leading-order ChPT cross section, normalized to the respective maximum of the distribution. The left plot is for $m_{3\pi} = 0.475\,\mathrm{GeV}/c^2$, the right one for $0.642\,\mathrm{GeV}/c^2$, *i.e.* at the lower and upper ends of the investigated region. The quantities $m_{13}^2$ and $m_{23}^2$ are the squared masses of the $\pi^-\pi^+$ subsystems.

contributions from diffractive excitation of the incoming pions to the $a_1(1260)$ and $\pi_2(1670)$ resonances are clearly seen. In the highlighted low-mass region, $m_{3\pi} < m_{3\pi}^{\mathrm{lim}} \equiv 0.72\,\mathrm{GeV}/c^2$, leading-order ChPT is expected to be applicable [7]. The momentum transfer distribution for these low masses, excluding the kaon peak, is depicted in Fig. 3. The sharp increase with $t' \to 0$ includes the Primakoff contribution that is investigated in detail in the following.

The data are analyzed using an extension of the partial-wave analysis (PWA) package that was previously used for the high-$t'$ events from the same data set [8]. The $X^- \to \pi^-\pi^-\pi^+$ transition is modeled as a two-step process with an intermediate $2\pi$-resonance (isobar), $X^- \to (2\pi\,\mathrm{isobar})^0\pi^- \to \pi^+\pi^-\pi^-$, specified as $J^{PC}M^{\varepsilon}(2\pi\,\mathrm{isobar})[L]\pi$. Here $J$ is the spin, $P$ the parity and $C$ the charge-conjugation parity of $X$, $M$ the projection of $J$ onto the beam axis, $\varepsilon$ the reflectivity of the final state in the Gottfried-Jackson frame (see e.g. [8]), and $L$ is the orbital angular momentum between the isobar and the bachelor pion. The partial waves found to be relevant at low $m_{3\pi}$ and low $t'$ are summarized in Table 1.

The new element of the present analysis is the replacement of the $M{=}\,1$ isobaric waves at low masses by a single wave that is given by the fully differential form of the ChPT prediction and called from now on "chiral amplitude", see Eq. (17) in Ref. [1]. Replacing only $M{=}\,1$ waves accommodates the transverse nature of the quasi-real photon as exchange particle, as $M{=}\,0$ is forbidden for real photons and strongly suppressed at small $t'$. This replacement is based on the observation [9] that $\pi^-$Pb strong interaction is characterized by a $(t')^M\exp(-b_{\mathrm{Pb}}t')$ behaviour at high energy and small $t'$, with the Pomeron slope parameter $b_{\mathrm{Pb}} \approx 400\,(\mathrm{GeV}/c)^{-2}$. Thus for strong $\pi^-$Pb interactions at lowest $t'$, $M{=}\,0$ components dominate and $M{=}\,1$ are suppressed, whereas photon exchange features a sharp spike at $t' \approx |t|_{\mathrm{min}} \leq 8\cdot 10^{-5}\mathrm{GeV}^2/c^2$ for masses $m_{3\pi} < m_{3\pi}^{\mathrm{lim}}$, due to the $t'$ dependence given by Eq. (1).

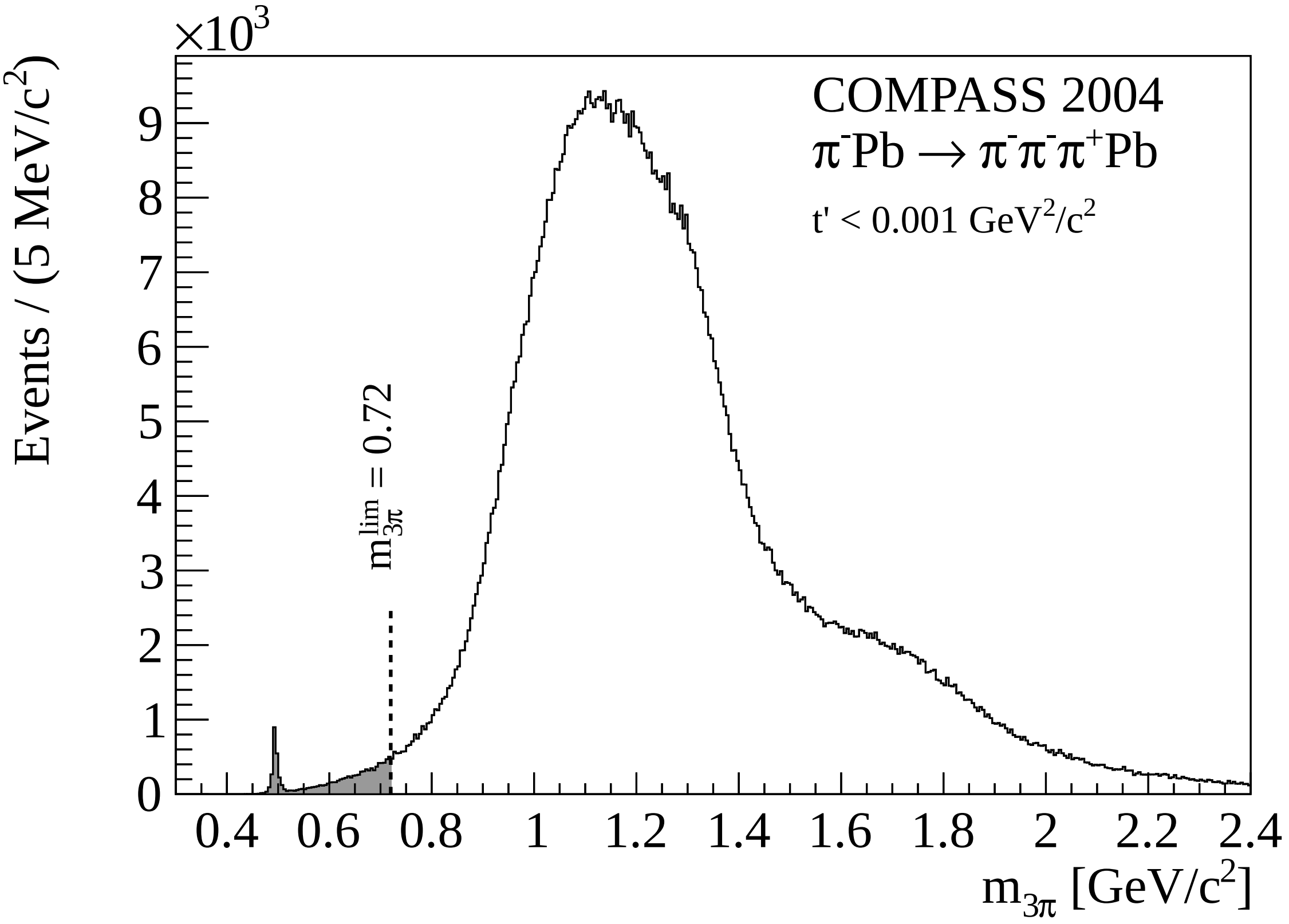

**Fig. 2:** Invariant mass spectrum of the final-state $3\pi$ system observed in the Primakoff region ($t' < 0.001\ \mathrm{GeV}^2/c^2$), with the interaction point observed in the target region. The kaon decay into $\pi^-\pi^-\pi^+$ is seen as sharp peak at $m_K \approx 0.493$ GeV. The low-mass region of interest for this Letter is highlighted.

**Table 1:** Wave sets used in the low-mass, low-$t'$ region. In the figures showing PWA fits, in the low-mass region exclusively the "chiral amplitude" is used. In addition to the indicated waves, always a background contribution that is homogeneous in phase space and the kaon decay contribution are allowed. Due to the very forward kinematics, for the $M$=1 waves both reflectivities are allowed, as discussed in the text.

| $M = 0$ | $M = 1$ | |
|---|---|---|
| $0^{-+}0^{+}(\pi\pi)_S[S]\pi$ | $1^{++}1^{\pm}(\pi\pi)_S[P]\pi$ | or { chiral amplitude |
| $0^{-+}0^{+}\rho[P]\pi$ | $1^{++}1^{\pm}\rho[S]\pi$ | |
| $1^{++}0^{+}\rho[S]\pi$ | $1^{-+}1^{\pm}\rho[P]\pi$ | |
| $1^{++}0^{+}(\pi\pi)_S[P]\pi$ | $2^{++}1^{\pm}\rho[D]\pi$ | |
| $2^{-+}0^{+}(\pi\pi)_S[D]\pi$ | $2^{-+}1^{\pm}\rho[P]\pi$ | |
| | $2^{-+}1^{\pm}(\pi\pi)_S[D]\pi$ | |

In order to study the mass dependence of the partial waves in the Primakoff region $t' < 0.001\,\mathrm{GeV}^2/c^2$, the low-mass ($m_{3\pi} < m_{3\pi}^{\lim}$) spectrum is subdivided into seven bins. In this region the PWA is performed with the isobar $M$= 0 waves of Table 1 and the chiral amplitude for $M$= 1. Above this limit both $M$= 0 and $M$= 1 isobaric wave sets of Table 1 are used. In order to study the exchange mechanisms in this low-mass region, we look at the $t'$ distribution extending the range up to $0.01\,\mathrm{GeV}^2/c^2$. The intensities are obtained from a PWA with the $M$= 0 set of Table 1 and the chiral amplitude for $M$= 1 both below and above $0.001\,\mathrm{GeV}^2/c^2$. The results of these analyses are shown in Figs. 4 and 5.

Figure 4 presents the fitted partial-wave intensities for the whole mass range, summed separately for $M$= 0 and $M$= 1. The intensities are corrected for acceptance effects that are determined by a Monte Carlo simulation and applied in the fitting process in their fully differential form in the 3-pion phase space. This correction does not deviate more than $\pm$ 8% from the average value of 0.58 over the kinematic range considered. About 82% of the total intensity is attributed to $M{=}0$ waves, which are dominated by the $a_1(1260)$ and $\pi_2(1670)$ resonances.

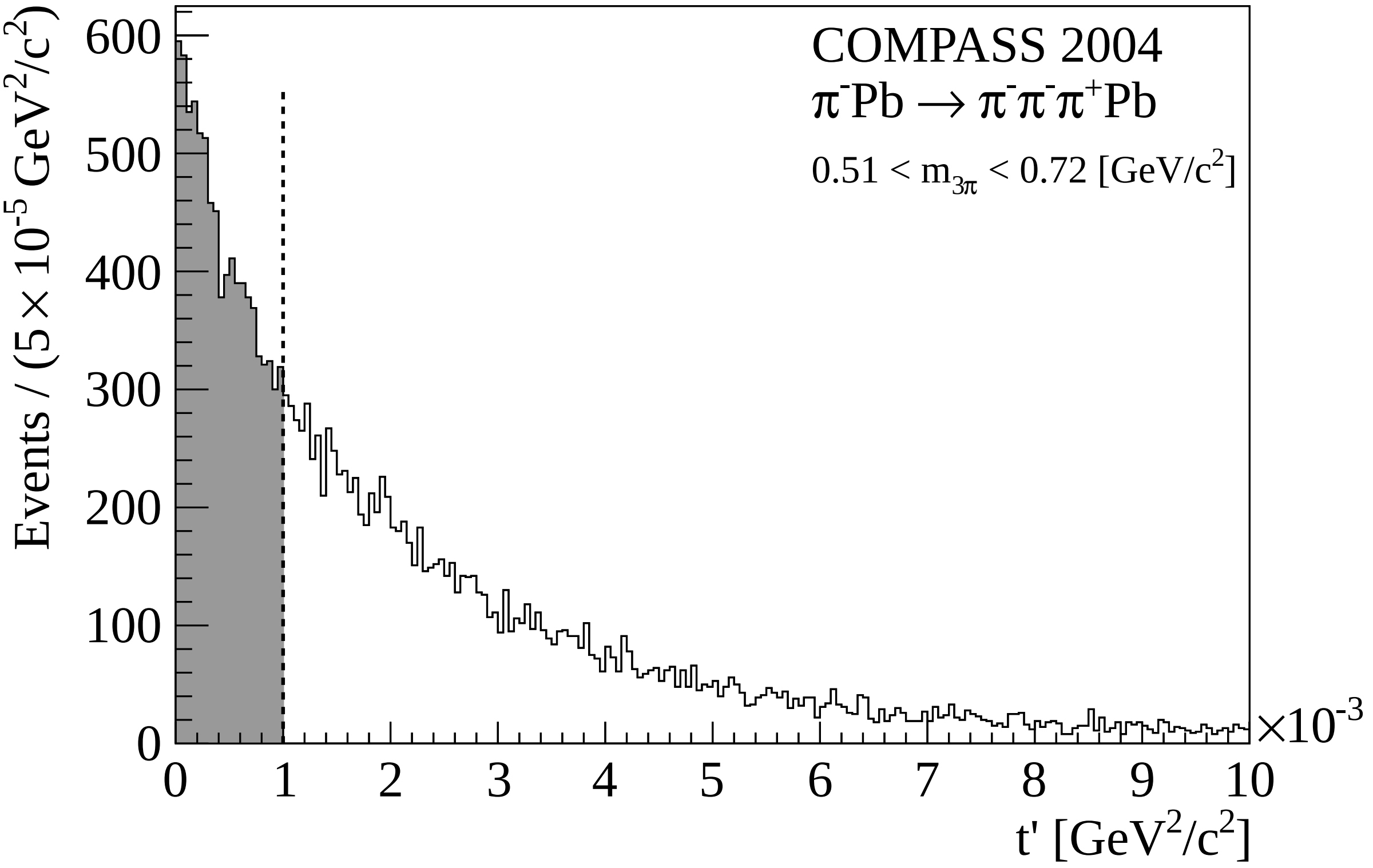

**Fig. 3:** Momentum transfer distribution of events in the investigated low-mass region. The Primakoff region ($t' < 0.001\,\mathrm{GeV}^2/c^2$) is highlighted. In order to suppress the kaon contribution, $m_{3\pi} > 0.51\,\mathrm{GeV}/c^2$ is required.

For the low-mass region integrated between $0.51{-}0.72\,\mathrm{GeV}/c^2$, Fig. 5 shows the PWA results for the $t'$ distributions separately for the $M{=}0$ and ChPT (*i.e.* $M{=}1$) contributions. In both cases, an exponential function is fitted to the spectrum. For $M{=}0$, Pomeron exchange with the slope $b_{\mathrm{Pb}}$ as specified above is confirmed. For $M{=}1$, the exact shape expected for photon exchange can not be resolved experimentally since it is distorted by multiple scattering and detector resolution. The resulting spectrum is well described by an $\exp(-b't')$ function, where the theoretically "forbidden" forward region at $t' \to 0$ is filled by multiple scattering. The slope fitted to the ChPT intensity shown in Fig. 5, $b' = 1447 \pm 196\,(\mathrm{GeV}/c)^{-2}$, is in good agreement with the simulation of pure photon exchange, $b'_{\mathrm{sim}} \approx 1600\,(\mathrm{GeV}/c)^{-2}$. Strong interaction $M{=}1$ exchange, on the other hand, would have a slowly rising $\propto t'\exp(-bt')$ distribution. From additional studies performed for the neighboring $t'$ region, $0.001 < t' < 0.01\,\mathrm{GeV}^2/c^2$, strong interaction $M{=}1$ contributions to the Primakoff region were estimated to be well below 5%.

The fitted chiral amplitude very efficiently replaces all $M{=}1$ waves, especially the dominant $1^{++}1^{+}$ contributions also used as isobaric waves in PWA fit attempts of the low-mass region, see Table 1. Fitting the chiral amplitude as $M{=}1$ contribution leads to smaller statistical uncertainties due to the smaller number of free parameters, while it features a signal strength compatible with that of the sum of the $M{=}1$ waves in the isobaric description. The PWA attributes a comparable likelihood to both cases, which reflects that the real event distribution in the 5-dimensional phase space follows equivalently well the distributions expected from the two ansätze. But, the observed enhancement of $M{=}1$ waves at small $t'$ lacks a theoretical explanation other than its chiral nature, and the comparison to the fit with isobaric waves is presented here only as a well-known starting point of the employed PWA.

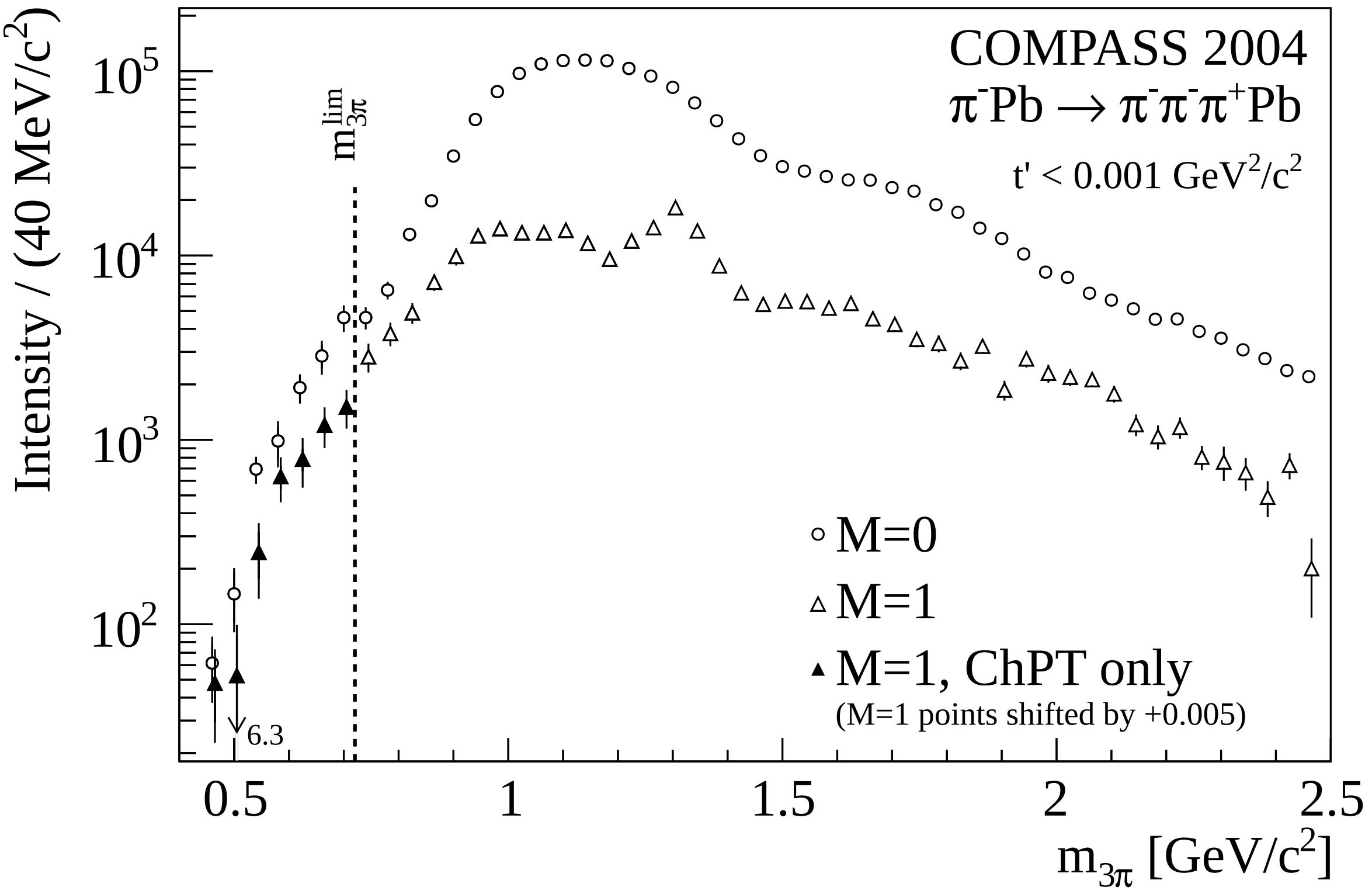

**Fig. 4:** PWA intensities summed separately for the $M = 0$ and $M = 1$ components. Total uncertainties are shown. Fluctuations of the systematic uncertainties, as discussed along with Fig. 6, were smoothed in the range $0.5 < m_{3\pi}/(\mathrm{GeV}/c^2) < 1$, and the obtained uncertainties were added quadratically to the statistical ones.

It is specific for the very forward kinematics studied here that the production plane spanned by the incoming beam and the outgoing $X^-$ system becomes less accurately defined as $t' \to 0$. As a consequence, the PWA not only shows the expected $M^\varepsilon = 1^+$ states, but also states with $M^\varepsilon = 1^-$ resulting from the limited resolution in the azimuthal orientation of the $3\pi$-system, which applies only for $M \neq 0$. As partial waves are also identified by their signature in the remaining 4 kinematical variables that are not distorted by resolution effects in the $t' \to 0$ limit, the two reflectivity contributions have been summed up and attributed to the well-motivated and expected $M^\varepsilon = 1^+$ intensities. The validity of this approach was confirmed by a special Monte Carlo simulation, where the appearance of negative-reflectivity contributions arising from purely positive-reflectivity input states through resolution effects of the set-up was confirmed.

In order to compare the observed $M = 1$ strength at $m_{3\pi} < m_{3\pi}^{\mathrm{lim}}$ with the ChPT cross section prediction, the intensities obtained from PWA must be normalized to incoming beam flux and target thickness. While the latter is known to high precision, the effective beam flux is influenced by a variety of effects which are less well controlled. The beam intensity was not monitored with high precision, and absolute trigger and detection efficiencies have to be modeled. The method employed here to overcome these uncertainties is based on the free decay of the kaon component in the beam [10]. As can be seen in Fig. 2, the observed 3-pion final state contains a contribution from kaon decays $K^- \to \pi^-\pi^-\pi^+$. Since the fraction of kaons in the beam and their decay kinematics are known to a much better precision than the statistical uncertainty, the observed kaon strength was used to determine the effective luminosity. Uncertainties in the Monte Carlo description of the experimental setup cancel out, since they affect the $K^- \to \pi^-\pi^-\pi^+$ decay and the $\pi^-\gamma \to \pi^-\pi^-\pi^+$ reaction in a similar way.

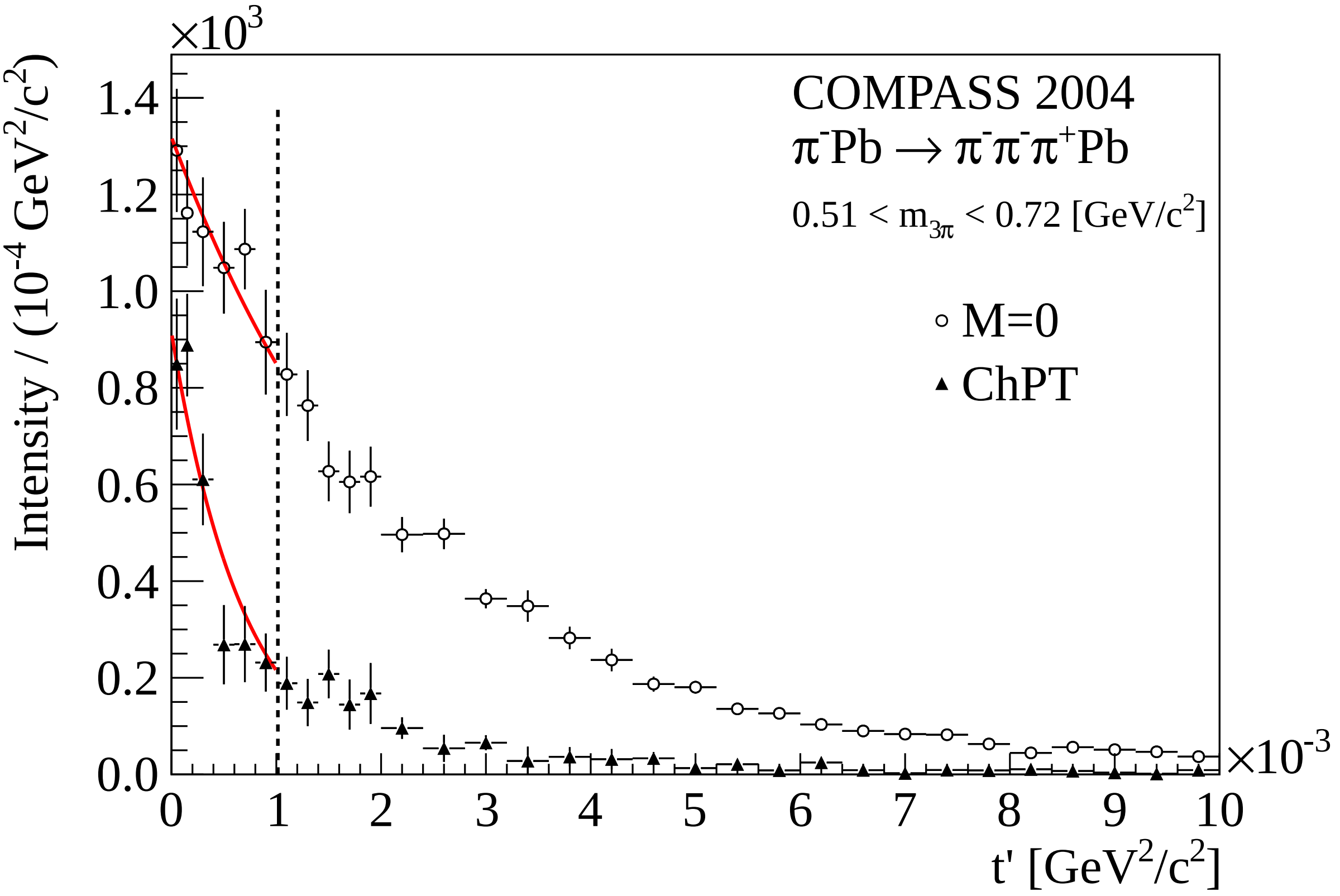

**Fig. 5:** PWA intensities for the investigated low-mass range. Each $t'$ bin represents an independent fit to the data, yielding simulaneously the two values for the $M=0$ and the chiral intensity.

In Fig. 6 the real-photon absolute cross section is shown. It has been obtained by taking into account the effective luminosity for the chiral amplitude and dividing out the quasi-real photon density factor as given in Eq. (1). A good agreement with the leading order ChPT prediction [1] is observed. This is expected since in ChPT loop effects and resonances are supposed to become important only at higher masses [7].

The systematic uncertainties shown in the figures contain the following components: (i) the variation due to the fitting model; as different fit models gave results with similar values for the likelihood, we attributed the corresponding variations of the wave intensities as systematic uncertainty; (ii) the uncertainty originating from the luminosity determination, which is also indicated separately in Fig. 6; (iii) the full calculation of the e.m. radiative corrections is not yet available for $\pi^-\gamma \rightarrow \pi^-\pi^-\pi^+$, in contrast to the case of $\pi^-\gamma \rightarrow \pi^-\pi^0\pi^0$. A first estimate indicates that this correction is well below 5% [11]. All these systematic uncertainties sum up to about 18% in total.

In conclusion we have measured the $\pi^-\gamma \rightarrow \pi^-\pi^-\pi^+$ cross section in the region $m_{3\pi} < 5m_\pi$, with a total uncertainty of about 20%. The measured points are in good agreement with the ChPT prediction. At this level of accuracy, ChPT therefore provides a good description of the strong pion-pion interactions at low energies, including the coupling to quasi-real photons via the Primakoff process.

We gratefully acknowledge the support of the CERN management and staff as well as the skills and efforts of the technicians of the collaborating institutions. The authors thank N. Kaiser for instructive and helpful discussions.

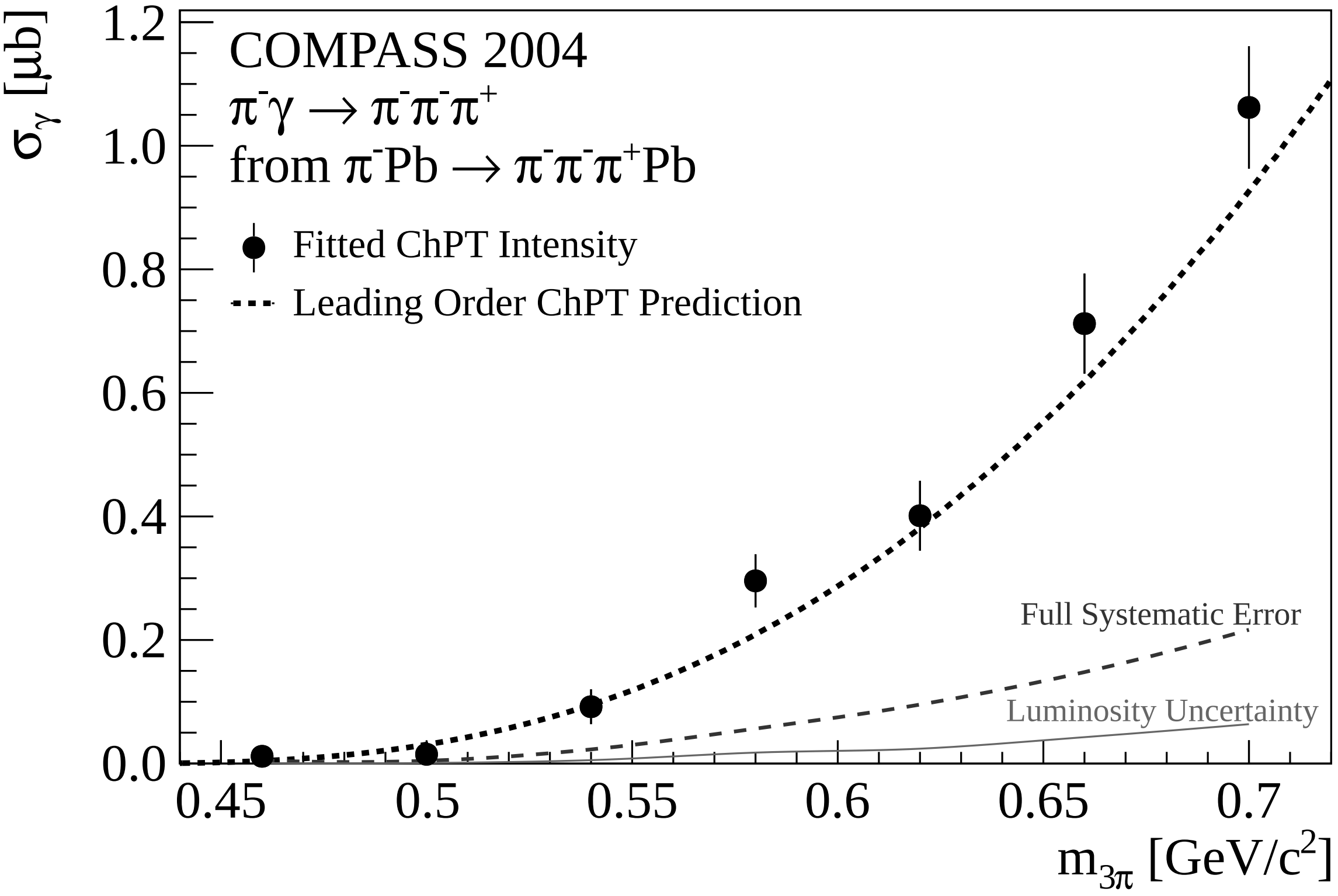

**Fig. 6:** Cross section for $\pi^-\gamma \to \pi^-\pi^-\pi^+$ as a function of the total collision energy $\sqrt{s} = m_{3\pi}$. The error bars show only the statistical error. The evolution of the systematic uncertainty, as discussed in the text and smoothed over the bins, is indicated as dashed line.